\documentclass[twocolumn]{aastex631}

\usepackage{amsmath}
\usepackage{xcolor}
\usepackage{hyperref}

\definecolor{blazeorange}{rgb}{1.0, 0.4, 0.0}
\definecolor{seagreen}{rgb}{0.18, 0.55, 0.34}
\definecolor{rufous}{rgb}{0.66, 0.11, 0.03}
\definecolor{royalfuchsia}{rgb}{0.79, 0.17, 0.57}
\definecolor{scarlet}{rgb}{1.0, 0.13, 0.0}
\definecolor{royalpurple}{rgb}{0.47, 0.32, 0.66}
\definecolor{darkblue}{rgb}{0, 0, 0.66}

\begin{document}

\title{A merger within a merger: Chandra pinpoints the short GRB 230906A in a peculiar environment}

\author[0000-0001-6849-1270]{Dichiara S.}
\affiliation{Department of Astronomy and Astrophysics, The Pennsylvania State University, 525 Davey Lab, University Park, PA 16802, USA}

\author[0000-0002-1869-7817]{Troja E.}
\affiliation{Department of Physics, University of Rome - Tor Vergata, via della Ricerca Scientifica 1, 00100 Rome, IT}
\affiliation{INAF, Viale del Parco Mellini 84, 00136 Roma, Italy}

\author[0000-0002-9700-0036]{O'Connor B.}
\affiliation{McWilliams Center for Cosmology and Astrophysics, Department of Physics, Carnegie Mellon University, Pittsburgh, PA 15213, USA}

\author[0000-0003-0691-6688]{Yang Y.-H.}
\affiliation{Department of Physics, University of Rome - Tor Vergata, via della Ricerca Scientifica 1, 00100 Rome, IT}

\author[0000-0001-7833-1043]{Beniamini P.}
\affiliation{Department of Natural Sciences, The Open University of Israel, P.O Box 808, Ra'anana 4353701, Israel}
\affiliation{Astrophysics Research Center of the Open university (ARCO), The Open University of Israel, P.O Box 808, Ra'anana 4353701, Israel}
\affiliation{Department of Physics, The George Washington University, 725 21st Street NW, Washington, DC 20052, USA}

\author[0000-0001-5193-3693]{Galvan-Gamez A.}
\affiliation{Instituto de Astronom\' ia, Universidad Nacional Aut\'onoma de M\'exico, Circuito Exterior, C.U., A. Postal 70-264, 04510 Cd. de M\'exico, M\'exico}

\author[0000-0001-6276-6616]{Sakamoto T.}
\affiliation{Department of Physical Sciences, Aoyama Gakuin University, 5-10-1 Fuchinobe, Chuo-ku, Sagamihara, Kanagawa 252-5258, Japan}

\author[0000-0002-2064-3164]{Kawakubo Y.}
\affiliation{Department of Physical Sciences, Aoyama Gakuin University, 5-10-1 Fuchinobe, Chuo-ku, Sagamihara, Kanagawa 252-5258, Japan}

\author[0000-0003-4877-9116]{Charlton J. C.}
\affiliation{Department of Astronomy and Astrophysics, The Pennsylvania State University, 525 Davey Lab, University Park, PA 16802, USA}




\begin{abstract}

We report the precise X-ray localization of GRB~230906A, a short duration ($T_{90}\sim$0.9 s) burst  with no optical or radio counterpart. 
Deep imaging with the \textit{Hubble Space Telescope} detects a faint galaxy (G$^\ast$; $F160W\simeq26$ AB mag) coincident with the sub-arcsecond X-ray position.
Compared with standard GRB galaxies, its faintness, compact size and color would suggest a high redshift ($z\gtrsim$3) host. 
However, our observations also reveal the presence of a galaxy group at $z\!\sim$0.453, confirmed spectroscopically with VLT/MUSE, with clear signs of interactions and mergers among group members. 
The GRB and its putative host project onto an extended ($\approx$180 kpc) tidal tail emerging from the group's central galaxy. The probability of a chance alignment is small ($P_{cc}\!\lesssim\!4$\%),
we thus argue that the GRB and its galaxy G$^*$
reside within the group.  Their peculiar location along the tidal debris
suggests that an enhanced burst of star formation, induced by the galaxy merger, might have formed the progenitor compact binary $\lesssim$700 Myr ago. 
The compact binary later evolved in a neutron star merger which produced GRB 230906A and injected $r$-process material into the surrounding circumgalactic medium.

\end{abstract}



\section{Introduction}
\label{sec:introduction}

Gamma-ray bursts (GRBs) separate into two broad classes by duration and spectral hardness \citep{Kouveliotou1993}: long/soft events linked to the collapse of massive stars, and short/hard bursts widely interpreted as the electromagnetic counterparts of compact object mergers \citep[e.g.,][]{Eichler1989, Nakar2007, Berger2014}. The gravitational wave transient GW170817 provided smoking-gun evidence connecting the gravitational wave signal of a binary neutron star (NS) merger with the short GRB 170817A and the kilonova AT2017gfo \citep{Abbott2017multi, Abbott2017grb, Arcavi2017, Coulter2017, Cowperthwaite2017, Drout2017, Kasliwal2017, Lipunov2017, Pian2017, Smartt2017, Troja2017, Valenti2017}. 

Having established the link between short GRBs and binary NS mergers, short GRB observations can be used as tracers of their progenitor population across cosmic time. 
Their redshift distribution, host-galaxy demographics, and galactocentric offsets jointly constrain the delay-time distribution (DTD) between star formation and merger, the typical magnitude of natal kicks imparted to compact binaries, and the role of mergers in the cosmic production of $r$-process elements \citep{Wanderman2015,Zevin2020,Beniamini2016}. 

On galactic scales, chemical evolution models that include NS mergers calibrated on the kilonova AT2017gfo and its inferred yields can reproduce Eu abundances \citep{Cote2018, Cowan2021}, and stellar archaeology in ultra-faint dwarfs demonstrates that a small number of prolific $r$-process events, such as NS mergers, can dominate a system’s enrichment history \citep{Ji2016}. 
Short GRBs statistics extend these constraints to higher redshift and a wider variety of environments than are currently accessible with gravitational-wave detections alone, enabling a joint picture in which afterglow/host observations calibrate DTDs and kick distributions while kilonova detections and upper limits inform nucleosynthetic yields. 

After two decades of \textit{Swift} operations \citep{Gehrels2004}, a large sample of well-localized short GRBs is now available to probe environments and offset distributions \citep{oconnor2022,Fong2022}. In practice, however, the subset with sub-arcsecond positions remains strongly optically driven, since optical detections provide the most precise localizations and hence the most robust GRB–galaxy associations. 
By contrast, X-ray positions from \textit{Swift}/XRT, even with astrometric enhancements \citep{Goad2007,Evans2009}, are typically accurate to a few arcseconds and multiple plausible hosts can reside within a single error region \citep{Gehrels2005, Bloom2006, Prochaska2006}. 
The dependence on optical detections introduces well-known selection effects, biasing samples toward lower redshift, less-extincted sightlines, and denser circumburst media where afterglows are intrinsically brighter \citep{oconnor2020}. 
Conversely, high-redshift short GRBs (where Lyman blanketing suppresses observed-frame optical bands) and mergers occurring in low-density or halo environments (which produce faint optical afterglows and large host offsets) may be underrepresented in the optical sample. Mitigating these biases requires sub-arcsecond localizations independent of optical brightness, which can be achieved in X-rays thanks to \textit{Chandra} imaging.

\begin{figure*}[!t]
\centering
\includegraphics[scale=0.70]{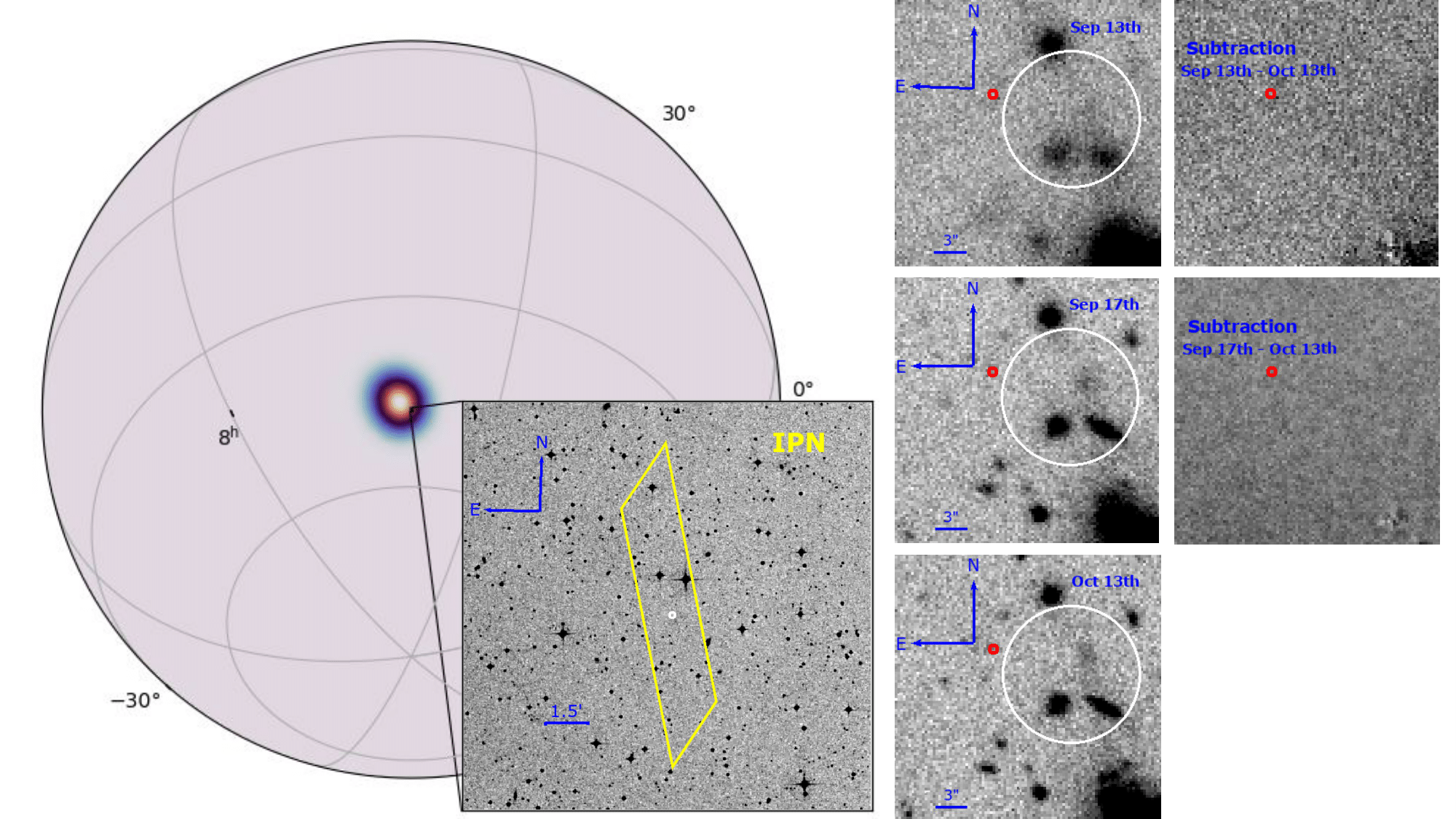}
 \caption{The color map of the initial {\it Fermi}/GBM localization (plotted using \texttt{ligo.skymap}) with the insert showing the IPN triangulation. On the left are the VLT images collected in the 3 epochs of observations together with the subtracted images. The red and white circles show the {\it Chandra} and XRT localization of the source, respectively
}
 \label{fig:loc}
\end{figure*}

In the absence of optical or radio counterparts, 
X-ray imaging with \textit{Chandra} proved essential for the precise localization of GRB 111020A \citep{Fong2012} and GRB 111117A \citep{Sakamoto2013}, two emblematic cases  representing the subset of short GRBs susceptible to optical selection effects. 
In the case of GRB 111020A, \textit{Chandra} pinpointed the X-ray afterglow well away from any bright galaxy. Deep optical and near-infrared follow-up revealed no plausible host to stringent limits, classifying the event as `hostless' \citep{Fong2012}.  
In the case of GRB~111117A, the \textit{Chandra} localization enabled a secure association with a faint galaxy at $z\approx2.21$, placing it among the most distant short GRB known to date \citep{Sakamoto2013,Selsing2018}.

Here we report a third case of a short GRB that, lacking optical and radio counterparts, was precisely localized in a peculiar environment through rapid \textit{Chandra} Target-of-Opportunity (ToO) observations. 
We leveraged the \textit{Chandra} imaging with late-time \textit{XMM-Newton} aimed at constraining the jet collimation, 
deep optical and nIR imaging carried out with the Very Large Telescope (VLT) and the \textit{Hubble Space Telescope}, and integral-field spectroscopy of nearby galaxies to map the redshift structure of the field.
The paper is organized as follows:
Details of the data reduction and analysis of the multi-wavelength observations of the short GRB 230906A are presented in Section~\ref{sec:observations}. We discuss the afterglow and the GRB environment properties in Section~\ref{sec:discussion}. The conclusions are summarized in Section~\ref{sec:summary}.

Uncertainties are quoted at the 68\% confidence level (1$\sigma$) for each parameter of interest and upper limits are given at a 3$\sigma$ level, unless otherwise stated.

\section{Observations and Data Analysis}
\label{sec:observations}

\subsection{Gamma-rays}
\label{sec:GBM}

On 2023 September 6th at 12:55:07.15 UT (hereafter $T_0$), the Gamma-ray Burst Monitor (GBM) aboard \textit{Fermi} triggered on GRB 230906A and localized it within a region of radius 2.8 degrees (statistical only) \citep{GCN34631}. 
The burst occurred $\approx$54 deg from the \textit{Fermi} LAT boresight and the source position remained within this boresight angle up to $T_0$+2183 s. The 90\% upper limit in LAT is 4.0$\times$10$^{-9}$  erg cm$^{-2}$ s$^{-1}$ (100 MeV - 100 GeV, assuming a photon index of 2.1). 

The burst was also detected by INTEGRAL (SPI-ACS), Konus-Wind, and Mars-Odyssey High-Energy Neutron Detector (HEND), which enabled triangulation of its position to a narrow error region measuring 12 arcmin in length and 2 arcmin in width (Figure~\ref{fig:loc}; \citealt{GCN34637}). This refined localization, derived via the InterPlanetary Network (IPN), proved essential for subsequent follow-up observations.

\begin{figure}[t]
\centering
\includegraphics[width=\columnwidth]{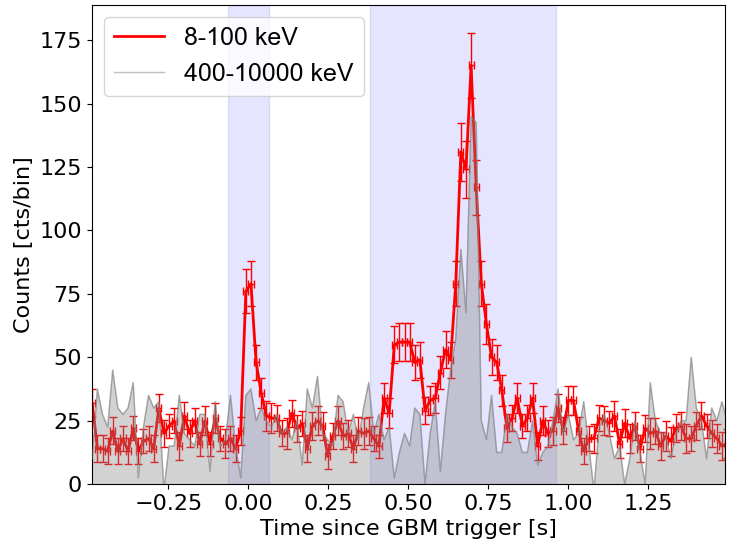}
 \caption{{\it Fermi}/GBM light curves of GRB 230609A in two energy bands, soft (8-100 keV) and hard (400-10000 keV). 
The time bin is 16 ms. Purple shaded regions highlight the first peak and the main episode
}
 \label{fig:lcurves}
\end{figure}

We downloaded the public \textit{Fermi} data from the archive\footnote{https://heasarc.gsfc.nasa.gov/FTP/fermi/data/gbm/triggers/} and analyzed them using the standard GBM tools v.2.0.8 \citep{2019ascl.soft05011F}. Spectra were fit using \textsc{RMFIT} v.4.3.2 \citep{2014ascl.soft09011G} by minimizing the Cash statistics. 
For the analysis we used the data collected by the two most illuminated NaI detectors (N1 and N2) and the B0 BGO detector.

The burst duration, defined as the time interval over which 90\% of the counts are measured, is $T_{90}$=0.94$\pm$0.35~s. This values is consistent with the one reported in the GBM catalog\footnote{https://heasarc.gsfc.nasa.gov/w3browse/fermi/fermigbrst.html}\citep{vonKienlin2020} and it was derived using the 16 ms light curve in the 8-1000 keV energy range, by modeling and subtracting the background with a linear function. The background was fitted over the time intervals T$_{0}$-4 - T$_{0}$-0.2 s before the signal, and T$_{0}$+2 - T$_{0}$+6.5 s after the signal.

We used the method presented in \citet{Kawakubo2015} to compute the spectral lag ($\tau$) using 16 ms light curves extracted in the 25–50 keV and 100–300 keV energy ranges \citep[following the procedure described in][]{Norris2000}. We found $\tau$=2.5$\pm$6.8 ms. This value is consistent with zero and aligns with previous observations of short GRBs \citep{Gehrels06}.

The gamma-ray light curve is formed of a first short and spectrally soft peak followed by a 0.3~s long period of apparent quiescence and a main episode (as shown in Figure~\ref{fig:lcurves}). 
The results of the time-resolved spectral analysis for the different intervals are presented in Table~\ref{tab:spectra}.
For any possible redshift, the time-averaged spectral properties (computed over the full duration of the burst) place GRB 230906A in the region of the Amati plot occupied by short GRBs \citep{Amati}.

\begin{deluxetable*}{lcccccc}
\tablecaption{Spectral analysis of the prompt emission \label{tab:spectra}}
\tablehead{\colhead{Interval} & \colhead{Time range} & \colhead{Fluence (10--1000 keV)} & \colhead{$E_{pk}$} & \colhead{$\alpha$} & \colhead{$\beta$} &   \colhead{C-STAT/}\\
 & s & $\times$10$^{-6}$ erg cm$^{-2}$ &  keV & index & index &  d.o.f}
\startdata
\hline
First peak & $T_{0}$-0.064--$T_{0}$+0.064 & 0.25$\pm$0.04 & 190$\pm$60 & $-$0.2$\pm$0.4 & $-$2.4$\pm$0.7 &  362/356 \\
Main peak & $T_{0}$+0.380--$T_{0}$+0.960 & 2.20$\pm$0.10 & 460$\pm$70 & $-$0.9$\pm$0.1 & $-$2.5$\pm$0.5 &  353/356 \\
Total duration & $T_{0}$-0.064--$T_{0}$+0.960 & 2.60$\pm$0.10 & 440$\pm$80 & $-$0.9$\pm$0.1 & $-$2.4$\pm$0.4 &  373/356 \\
\enddata
\tablecomments{Best fit results obtained over different time intervals. A Band function \citep{Band1993} was used to fit the spectra.}
\end{deluxetable*}

\subsection{X-ray Observations}
\label{sec:XRAY}

\begin{table*}
 	\centering
 	\caption{X-ray and Optical/nIR observations of GRB~230906A.
 	}
 	\label{tab: observations}
 	\begin{tabular}{lcccccc}
    \hline
    \hline
   \multicolumn{7}{c}{\textbf{X-ray}}  \\
   \hline
   Start Date  & Mid-time from T$_{0}$ & Exposure & Telescope & Instrument &  & Unabsobed Flux\\
    UT & d & ks & &  &  & 10$^{-14}$ erg cm$^{-2}$ s$^{-1}$\\
 \hline
2023-09-07 14:49:33 &  1.15 & 4.7 & {\it Swift} & XRT &  & 9.7$^{+5.0}_{-3.8}$\\
2023-09-11 04:59:37 &  4.67 & 19.8 & {\it Chandra} & ACIS-S & & 1.6$^{+0.5}_{-0.4}$\\
2023-09-17 15:02:29 &  11.46 & 47.0 & {\it XMM-Newton} & pn &  & 0.5$^{+0.1}_{-0.2}$\\
 \hline
 \hline
    \multicolumn{7}{c}{\textbf{Optical/nIR Photometry}}  \\
       \hline
    Start Date  & Mid-time from T$_{0}$ & Exposure & Telescope & Instrument & Filter & AB Mag\\
    UT & d & ks & &  &  &  \\
    \hline
   2023-09-07 14:50:04 & 1.15 & 4.7 & \textit{Swift} & UVOT & \textit{wh} & $>$23.0\\ 
   2023-09-13 07:09:10 & 6.79 & 4.8 & VLT & FORS2 & \textit{R} & $>$24.1$^{a}$\\  
   2023-09-13 08:12:26 & 6.83 & 5.5 & VLT & HAWK-I & \textit{Ks} & $>$22.2\\ 
   2023-09-17 07:06:17 & 10.78 & 4.8 & VLT & FORS2 & \textit{R} & $>$25.0$^{a}$\\
    \hline
     \hline
    \multicolumn{7}{c}{\textbf{Host Galaxy}}  \\
       \hline
    Start Date  & Mid-time from T$_{0}$ & Exposure & Telescope & Instrument & Filter & AB Mag$^{b}$\\
    UT & d & ks & &  &  &  \\
    \hline
   2023-10-13 06:18:00 & 36.75 & 4.8 & VLT & FORS2 & \textit{R} & 26.3$\pm$0.3\\ 
   2024-02-18 10:07:27 & 164.89 & 2.4 & \textit{HST} & WFC3 & \textit{F160W} & 26.0$\pm$0.1\\       
 \hline
    \end{tabular}
\begin{flushleft}
    \quad \footnotesize{X-ray fluxes are corrected for the Galactic absorption of $N_H = 2.45 \times 10^{20} \, \text{cm}^{-2}$ \citep{Willingale2013MNRAS.431..394W}. Optical/nIR upper limits correspond to a 3$\sigma$ confidence level, corrected for Galactic extinction \citep{Schlafly2011}}\\
    \quad \footnotesize{$^{a}$ Derived on the residual image, after the sbtraction}\\
    \quad \footnotesize{$^{b}$ Host Galaxy magnitude}
\end{flushleft}
\end{table*}

\subsubsection{Localization}
The {\it Neil Gehrels Swift Observatory} X-ray Telescope (XRT; \citealt{Burrows05}) began observing IPN localization at 14:46 UT, on September 7th, 2023 ($T_0$+1.8~d) for a total exposure of 4.7 ks. 
The X-ray afterglow was identified at the position RA, Dec. (J2000) = 5$^{h}$19$^{m}$01$^{s}$.79, $-$47$^{\circ}$ 53$^{\prime}$ 34$^{\prime\prime}$.9 with an uncertainty of 6.4 arcsec (90\% confidence level; \citealt{GCN34639}).
Due to the source faintness, an XRT enhanced position could not be derived. 
Table~\ref{tab: observations} reports the measured X-ray flux along with the observational details.

As shown in Figure~\ref{fig:loc}, several extended optical sources are visible  within the XRT localization, preventing a secure identification of the GRB host galaxy. For this reason, we triggered our approved program on the {\it Chandra}  X-ray Observatory (ID: 24500202; PI: Dichiara) aimed at achieving a sub-arcsecond localization. 

{\it Chandra} observations of GRB~230906A were carried out using the ACIS-S camera, starting at 4:58 UT on September 11th, 2023 ($\sim$\,$T_0$+4.7 d) for a total exposure of 19.1 ks.
We used the {\it Chandra} Interactive Analysis of Observations \citep[\textsc{CIAO} v. 4.14;][]{Fruscione2006} and the calibration files (CALDB 4.9.6) to reprocess the data in the 0.5-7.0 keV energy band.
The X-ray counterpart is detected with a high significance of 6.7$\sigma$ and a net count rate of 7.5$\times$10$^{-3}$\,cts\,s$^{-1}$, estimated using a circular extraction region of radius 1.5 arcsec. The background level was derived from a nearby source-free annulus region with internal and external radius of 3 arcsec and 7 arcsec, respectively.

A preliminary position, based on the {\it Chandra} native astrometry, was reported via the GRB Circular Network \citep{GCN34672}. 
We improved upon our previous work by aligning the X-ray image with the Legacy Surveys DR10 catalog \citep[][]{Dey2019} using the tool \texttt{wcs\_match} and 4 common sources.
Our refined position is RA, Dec. (J2000) = 05$^{h}$19$^{m}$01$^{s}$.53, -47$^{\circ}$53$^{\prime}$32$^{\prime\prime}$.4 with an uncertainty of 0.24\arcsec\ (68\% confidence level). 
This position represents the best localization of GRB~230906A. It lies slightly outside the XRT 90\% error region, but is within its 3\,$\sigma$ confidence level.

\subsubsection{Spectral Analysis}\label{sec:spec}
 
 To continue our monitoring of the X-ray afterglow, 
 we triggered our approved Target of Opportunity program (ID: 088306; PI: Troja) on 
 \textit{XMM-Newton}. 
 Observations were carried out on September 17, 2023 ($T_0$+10.9 d) using the EPIC cameras \citep{Struder2001,Truner2001} in Full window mode with the thin optical-blocking filter for approximately 60 ks. 
 Data were processed using the Science Analysis System (\textsc{SAS} v21.0.0) software and the latest calibration files. After standard filtering and removal of high background periods, the resulting exposures were 55.9 ks, 57.2 ks, and 47 ks for the MOS1, MOS2, and pn camera, respectively. 
 
 To minimize contamination from nearby X-ray sources, we used a circular source extraction region with a radius of 11 arcsec. 
 The background level was estimated from a nearby circular region of  50 arcsec radius for the pn image, and from a concentric source-free annulus for the MOS images. 
 The source is well detected ($\sim$4\,$\sigma$) in the pn image with a net count rate of (1.4$\pm$0.2)$\times$10$^{-3}$\,
 cts s$^{-1}$ (0.3--10 keV), derived after correcting for PSF-losses. 
 It is instead only marginally visible ($\sim$2.2\,$\sigma$)  in the combined MOS images, which are therefore not included in our spectral analysis. 
 
We grouped the source spectrum to a minimum of 1 count per bin and the background spectrum to a minimum of 3 counts per bin. The response matrix and ancillary response files were generated using the \texttt{rmfgen} and \texttt{arfgen} tasks.
We used \textsc{XSPEC} v12.14  \citep{Arnaud96} to fit the pn spectrum with an absorbed power-law function. 
The best fit model, obtained by minimizing the W statistics, has a photon index $\Gamma$ = 1.3$\pm$0.2 and an absorption $N_H = 2.45 \times 10^{20} \, \text{cm}^{-2}$, fixed to the Galactic value \citep{Willingale2013MNRAS.431..394W}. 
This model yields an unabsorbed flux of $6.3^{+2.1}_{-1.7} \times 10^{-15} \, \text{erg} \, \text{cm}^{-2} \, \text{s}^{-1}$ in the 0.3-10 keV band.

\begin{figure}[t]
\centering
\includegraphics[width=0.47\textwidth]{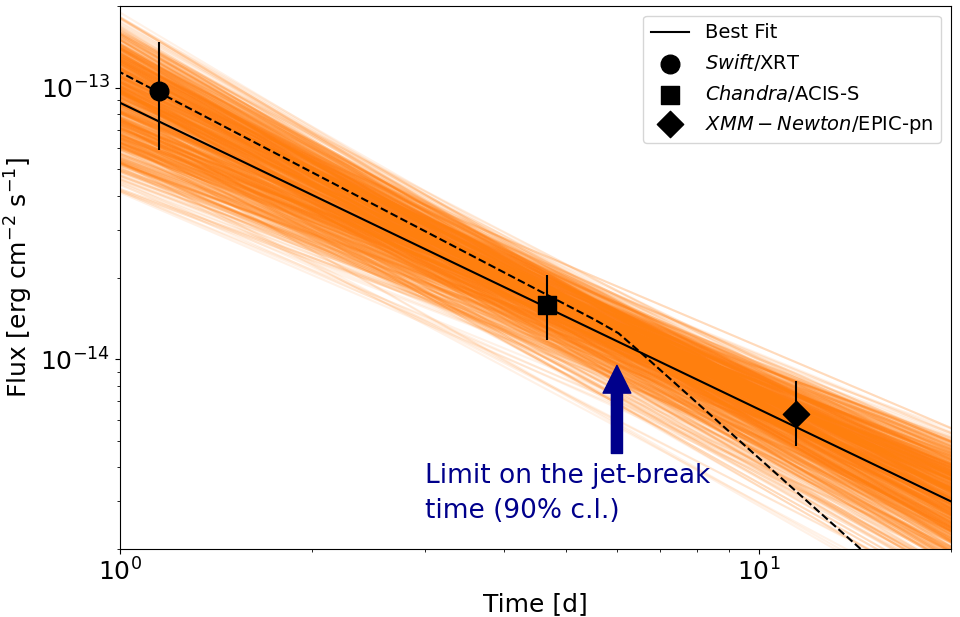}
 \caption{X-ray light curve, including {\it Swift}/XRT, {\it Chandra} and {\it XMM} data. The fluxes are corrected for Galactic absorption $N_H$ = 2.45 $\times$ 10$^{20}$ cm$^{-2}$ \citep{Willingale2013MNRAS.431..394W}. Orange lines show the 1 sigma range (derived from the MCMC fit) around the best for model, obtained assuming a power-law decay. The dashed line shows the best fit obtained assuming a broken power-law with a break at $T_{0}$+6 d, and a post-break decay index of 2.1 (the 90\% lower limit. See Section 2.2.3). 
}
 \label{fig:xraylc}
\end{figure}

\subsubsection{Temporal Analysis}

To study the temporal evolution of the X-ray counterpart, we calibrated the \textit{Swift}/XRT, \textit{Chandra}/ACIS-S and \textit{XMM}/EPIC observations using the same spectral model (Sect.~\ref{sec:spec}). 
The resulting fluxes are reported in Table~\ref{tab: observations}. 
Data were fit using the Markov Chain Monte Carlo (MCMC) algorithm implemented in the Python package \texttt{emcee}\citep{emcee}.
The combined light curve displays a simple power-law decay, $F_{\nu} \propto t^{-\alpha}$, with slope $\alpha$=1.13$_{-0.27}^{+0.35}$ (see Figure~\ref{fig:xraylc}). 

We attempted to constrain the presence of a jet-break by forcing in the fit a temporal break followed by a steep decline with $\alpha_2$=2.1. 
We then used the 90th percentile of the posterior distribution of the break time to place a 90\% lower limit  $t_j >6$ ~d to the jet-break time.

\subsection{Optical and nIR observations}
\label{sec:VLT}

\subsubsection{VLT/FORS2 imaging}
The FOcal Reducer and low dispersion Spectrograph \citep[FORS2; ][]{Appenzeller1998} mounted on the VLT started observing the field of GRB 230906A on September 13, 2023 at 7:09 UT ($T_0$+6.76, Program ID: 110.24CF.011, PI: Tanvir). 
Imaging consisted of 16$\times$300 s exposures in the $R$ filter, 
taken at an average airmass of 1.35 and in poor observing conditions
(seeing$\sim$2.6$^{\prime\prime}$). 
The same observations were repeated in better conditions on September 17 (seeing$\sim$1.1$^{\prime\prime}$) and October 13 (seeing$\sim$0.8$^{\prime\prime}$), $T_0$+10.76 d and $T_0$+36.72 d, respectively. 
Near-infrared (nIR) observations were collected with the High Acuity Wide field K-band Imager (HAWK-I; \citealt{HAWKI}). A sequence of 30$\times$60 s exposures was obtained in the $K$-band starting on September 13, at 8:12 UT
($T_0$+6.80). 

Preliminary results of these observations were reported via the GRB Circular Network by \citet{GCN34704}, who also claimed the possible detection of an optical counterpart. 
We downloaded the public data from the ESO archive
\footnote{http://archive.eso.org/eso/eso\_archive\_main.html} 
and reduced them 
using standard tools and CCD reduction techniques (e.g., bias subtraction, flat-fielding, etc.). 
Astrometry correction was applied to the single frames and they were aligned and stacked using \textsc{SCAMP} \citep{SCAMP} and \textsc{SWarp} \citep{Swarp}, respectively.

Using the images collected on October 13, at the sub-arcsecond position of the X-ray afterglow we detect a faint source with $R$=26.3$\pm$0.3 AB mag \citep[not corrected for the galactic extinction of E(B-V) = 0.020;][]{Schlafly2011}.
To assess the presence of an optical afterglow, we performed image subtraction using the latest imaging epoch as template (see Figure~\ref{fig:loc}). 
We performed two independent analyses using the High Order Transform of PSF ANd Template Subtraction (\textsc{HOTPANTS}) \citep{Becker2015} and the saccadic fast Fourier transform (\textsc{SFFT}) algorithm \citep{SFFT}, respectively. 
Our analysis does not reveal any significant residual within the \textit{Chandra} position, and sets a 3\,$\sigma$ upper limits of $R$\,$>$24.2 AB mag and $K$\,$>$22.2 AB mag to any optical and nIR counterpart at $T_0$+6.8 d and a limit of $R$\,$>$25.0 AB mag at $T_0$+10.8 d. 

\begin{figure*}
\centering
\includegraphics[scale=0.50]{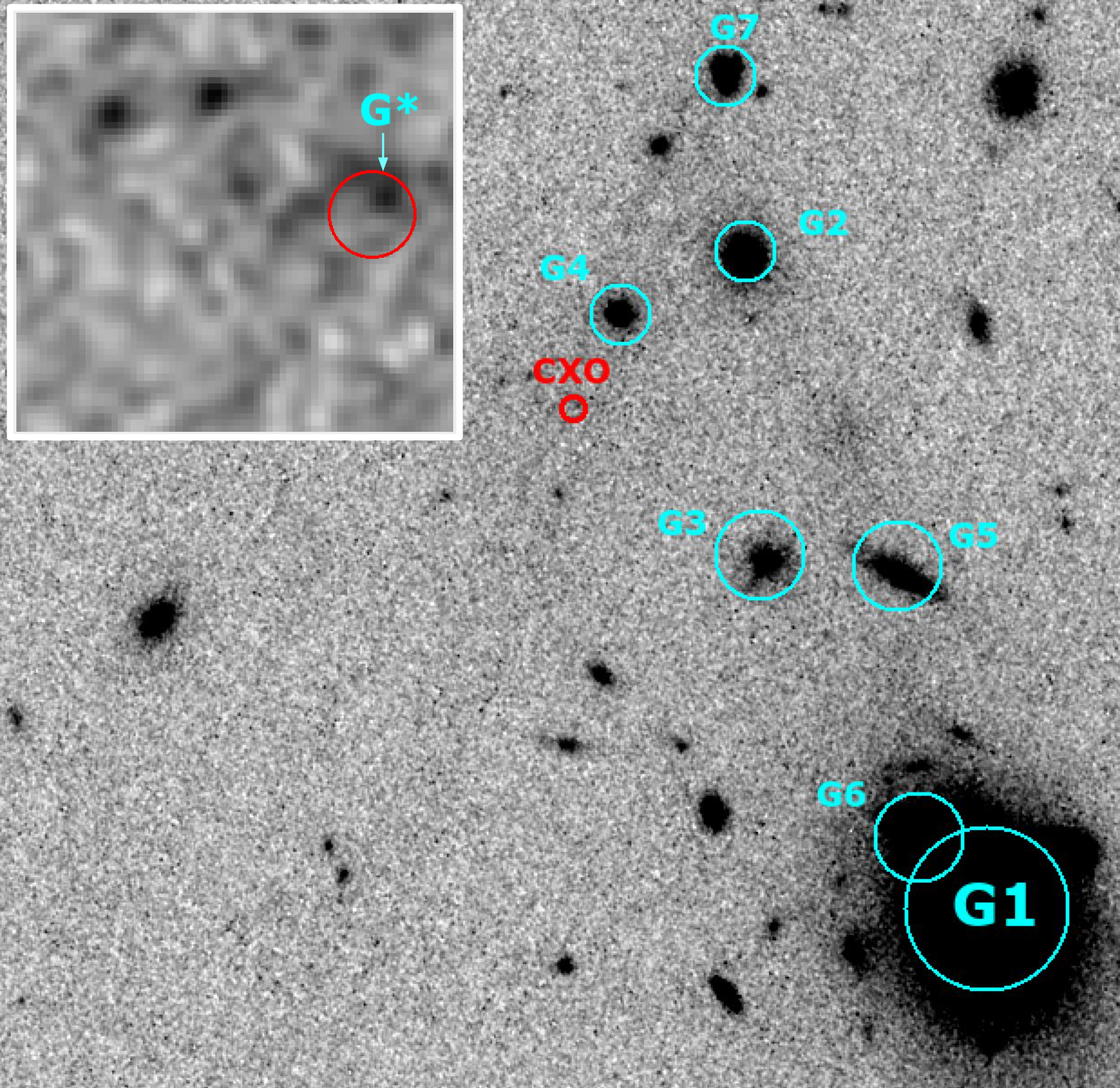}
 \caption{HST $F160W$ image of the GRB 230906 field. 
 The \textit{Chandra} localization region is marked by the red cyrcle.
 A zoom on the putative host galaxy (marked as G$^{*}$, with $P_{cc}$=4\%) is shown in the upper-left insert.
 Other nearby galaxies are numbered in ascending order of their $P_{cc}$ values, being G1 the next most probable one with $P_{cc}$=14\%. 
Galaxies that are not marked are characterized by $P_{cc}$ $>$ 70\%. 
}
 \label{fig:pcc}
\end{figure*}

\subsubsection{Hubble Space Telescope (HST)}

\textit{HST} observed GRB 230906A on February 18, 2024 (Program ID: 17492, PI: O'Connor)
using the Wide Field Camera 3 (WFC3) with the $F160W$ filter. 

The data were processed in a standard fashion using the \textsc{DrizzlePac} package \citep{Gonzaga12} to align, drizzle, and combine exposures. Frames were aligned to a common world-coordinate system using the \texttt{TweakReg} package. Next, the \texttt{AstroDrizzle} software was used to eliminate cosmic rays and bad pixels, generating the final combined drizzled image with pixel scale of 0.06\arcsec/pixel and a pixfrac value of 0.8. Photometric zero-points were derived from the keywords in the image header. 

The GRB field is shown in Figure~\ref{fig:pcc}. Within the X-ray localization, a nIR source, marked as $G^*$, is well detected with $F160W$= 26.00 $\pm$0.10 AB mag (see inset in Figure~\ref{fig:pcc}).
At these faint magnitudes, galaxies outnumber field stars by a large margin. 
Moreover, the \texttt{CLASS\_STAR} classifier from Source Extractor \citep{Bertin96} estimates a low probability that the nIR source is point-like (\texttt{CLASS\_STAR}=0.35). 
However, this classifier is optimized to work best on ground-based images with objects of moderate magnitude ($R<$23 mag). In our case, its results require further validation. 
Therefore, we adopted the star-galaxy separation criterion derived from \textit{HST} observations of the Great Observatories Origins Deep Survey (GOODS)
South field \citep{Windhorst11}. 
This technique correlates an object’s magnitude with its full-width at half-maximum (FWHM), allowing for a clear separation of the stellar locus from galaxies down to approximately 25.5–26 AB mag.
The location of $G^*$ in the magnitude/FWHM plot confirms that it is likely a faint galaxy. 

\subsubsection{VLT/MUSE spectroscopy}

The imaging observations presented in Sections 2.3.1 and 2.3.2 reveal the presence of several extended objects with similar colors, suggesting a possible overdensity of galaxies at a common redshift.  
To further investigate this possibility, we 
used the Multi Unit Spectroscopic Explorer \citep[MUSE;][]{2010SPIE.7735E..08B}, mounted on UT4 at the Very Large Telescope (VLT) on Cerro Paranal (Chile), to observe the field of GRB 230906A (Program ID: 114.27LW.025, PI: Troja). The instrument has a field of view of $1^{\prime} \times 1^{\prime}$. Observations were conducted with a seeing of 0.9 arcsec in Wide Field Mode (WFM-NOAO-N), i.e., without adaptive optics and using the nominal wavelength range. This configuration provides a spatial resolution of $0.2^{\prime\prime}$ per pixel and covers a spectral range of approximately 4700–9300~\AA. The spectral resolving power varies across this range, from $R \approx 1770$ at 4800~\AA\ to $R \approx 3590$ at 9300~\AA.

We obtained four exposures of 700\,s each.
Data were reduced using \textsc{ESO-Reflex} \citep[version 2.11.5;][]{2013A&A...559A..96F} and the MUSE pipeline package (version 2.10.10). The four exposures were then combined using \textsc{IRAF} \citep[v2.16;][]{1986SPIE..627..733T}. Spectra were extracted for 25 galaxies in the field, and secure redshift measurements were obtained for 18 of them, as listed in Table~\ref{tab:musespec}.

\section{Discussion}
\label{sec:discussion}

\subsection{Afterglow properties}

We interpret X-ray emission as synchrotron radiation produced by a forward shock expanding into a constant-density environment \citep{meszarosrees, wijersgalama, sari98, granot2002}. 
In a slow cooling regime ($\nu_m<\nu_c$), the spectral slope derived from the XMM spectrum ($\beta_X$ = 0.3$\pm$ 0.2) is consistent with standard closure relations if the cooling frequency ($\nu_{c}$) is above the X-ray band. A lower value of the cooling frequency would imply a spectral index, $p$, lower than 1
for the accelerated electrons \citep{granot2002,Gao13}. 
If the X-ray band is between the synchrotron frequency $\nu_{m}$ and $\nu_c$, the spectral slope $\beta_X$ implies an index $p\!=\!2\,\beta_X\!+\!1\!=\!1.6\!\pm\!0.4$  and the temporal decay index $\alpha_X\sim1.1$ implies $p\!=\!(4\,\alpha\!+\!3)\!/3\!=\!2.5^{+0.5}_{-0.4}$. Closure relations are thus satisfied for $p$\,$\sim$2.1-2.2.

In comparison to other short GRBs, the properties of the X-ray counterpart appear fairly standard. Assuming a decay rate $\alpha_X\sim1.1$, we estimate the X-ray afterglow flux at 11 h, $F_{X,11}$, and compare it to the observed gamma ray fluence in the 15-150 keV band, $f_{\gamma,15-150}$. 

The resulting ratio is log($F_{X,11}$/$f_{\gamma,15-150}$)=$-$6.5$_{-0.5}^{+0.3}$.  
This is consistent with the values observed for other short GRBs \citep[e.g.][]{oconnor2022,oconnor2020} for a wide range of projected physical offsets, and does not point to a particularly rarefied environment.

Assuming that the optical and X-ray bands belong to the same regime $\nu_m<\nu_o<\nu_X<\nu_c$ as $p\sim2.2$, we use a fiducial value of $\beta_{OX}$=0.6 to extrapolate the observed X-ray flux at optical wavelengths. We estimate the afterglow flux at $T_{0}$+1.15 d to be 25.9 AB mag in the UVOT $wh$ filter. This value is well below our limit and beyond the reach of many ground-based facilities. Only rapid ($\lesssim$ 24 h) observations with large aperture telescopes might have been able to detect the optical counterpart in favorable weather conditions. 
At later times ($T_{0}$+6.8 d) the predicted flux is 27.8 and 26.9 AB mag in the $R$-band and $K_s$-band, respectively, which was too faint for detection. 

Whereas the deep optical limits derived for GRB 111117A ($r \gtrsim$25.8 AB mag at 7.9 hr; \citealt{Sakamoto2013}) revealed a relatively low optical-to-X-ray flux ratio ($F_R / F_X \lesssim$590), possibly indicating a dusty sightline, the lack of an optical detection for GRB 230906A is not particularly constraining and can be attributed to a combination of weak afterglow and unfavorable observing conditions.

\begin{figure}[t]
\centering
\includegraphics[width=0.47\textwidth]{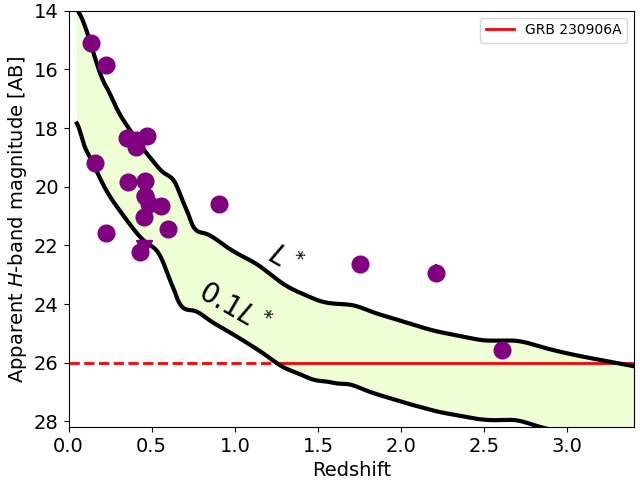}
 \caption{Observed H-band magnitude of a sample of short GRB host galaxies with spectroscopic redshifts \citep{Fong13, oconnor2022}. The black line contains the 0.1$L^*$-$L^*$ range of the evolving luminosity function \citep{Chen2003,Poli2003,Saracco2006,Marchesini2007,Marchesini2012,Hill2010,Ramos2011,Stefanon2013}. The red line marks the magnitude of the putative host galaxy G*, 
 which fits within the typical population of short GRB galaxies only for $z\gtrsim$1.2. 
}
 \label{fig:galaxies}
\end{figure}

\begin{figure}[t]
\centering
\includegraphics[width=0.47\textwidth]{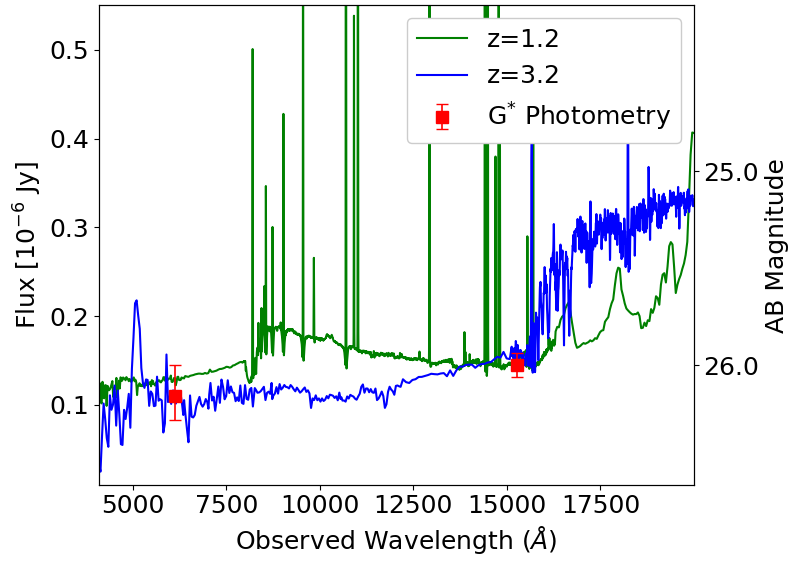}
 \caption{Spectral energy distribution of G$^*$ compared with two possible models: a low-mass ($\approx2\times10^8 M_{\odot}$) actively star-forming galaxy at $z\sim$1.2 (green) and a massive ($\approx9\times 10^9 M_{\odot}$) galaxy at $z\sim$3.2 (blue). 
}
\label{fig:sedg}
\end{figure}

\subsection{Environment}

\subsubsection{Association with $G^*$ and high-redshift origin}

The precise localization of GRB 230906A by the \textit{Chandra} X-ray Observatory enables a detailed investigation of its immediate environment. Within the \textit{Chandra} error circle, a faint galaxy coincident with the GRB position is detected by \textit{HST} and VLT. Therefore, we first examine the possibility that this is the GRB host galaxy. 

We followed the standard method outlined in \citet{Bloom2002} to estimate the probability of a chance alignment between  G$^*$ and the \textit{Chandra} position. 
We  used the updated formulation of \citet{Becerra23}, which is based on the $r$-band number counts from deep optical imaging \citep{Metcalfe2001,McCracken2003,Kashikawa2004}, and derived a density $\Sigma$\,$\sim$\,0.026 arcsec$^2$ of galaxies brighter than 26.3 AB mag. The resulting chance coincidence probability is 
$P_{cc} = 1 - {\rm exp}\left(-\pi r_X^2 \Sigma \right)$\,$\approx$4\%,
where $r_X$\,=\,0.7\arcsec\ is 
the 3\,$\sigma$ error radius of the X-ray position. 

If G$^*$ belongs to the population of short GRB host galaxies, it would be among the faintest detected to date
with an apparent infrared magnitude of $F160W \approx 26$ AB mag. To be consistent with the known host luminosity distribution, its faintness implies a redshift $z \gtrsim 1.2$ (see Figure~\ref{fig:galaxies}). 
The flat spectral slope observed between the optical and near-infrared bands suggests a relatively smooth continuum, with no prominent Balmer break detected. This absence may indicate either a young stellar population with intrinsically weak Balmer/4000 Å features \citep[e.g.][]{PoggiantiBarbaro1997,Gallazzi2005}, not common among short-GRB host galaxies, or a higher redshift ($z\gtrsim3$) at which the Balmer break is redshifted beyond the wavelength range covered by our data \citep[Figure~\ref{fig:sedg}. See also][]{Nanayakkara2024, Fontana2006}.

In the latter case, GRB 230906A could be one of the most distant short GRB ever discovered. 

\begin{table}
 	\centering
 	\caption{Redshift, chance coincidence probability and offsets derived for the field galaxies of GRB 230906A
 	}
 	\label{tab:gal}
 	\begin{tabular}{lccccc}
    \hline
   Name  & photo-z$^{a}$ & spec-z$^{b}$ & $P_{cc}$ & Ang. off. & Phys. off.$^{c}$\\
    & & & & arcsec & kpc\\
 \hline
G$^{*}$ &  - & - & 0.04 & - & -\\
G1 &  0.58 & 0.453 & 0.14 & 21.7 & 129\\
G2 &  0.49 & 0.453 & 0.26 & 7.6 & 44\\
G3 &  0.63 & 0.737 & 0.42 & 8.1 & 61\\
G4 &  - & - & 0.46 & 3.5 & -\\
G5 &  0.48 & 0.438 & 0.59 & 12.2 & 71\\
G6 &  0.45 & 0.451 & 0.66 & 18.3 & 109\\
G7 &  0.74 & 0.852 & 0.69 & 12.2 & 96\\
 \hline
 \hline
    \end{tabular}
\begin{flushleft}
    \quad \footnotesize{$^{a}$ From Legacy Survey DR10. G4 has no available estimate for either photometric or spectroscopic redshift.}\\
    \quad \footnotesize{$^{b}$ From MUSE observations. See Section 2.3.2}\\
\end{flushleft}
\end{table}

\begin{figure}
\centering
\includegraphics[scale=0.40]{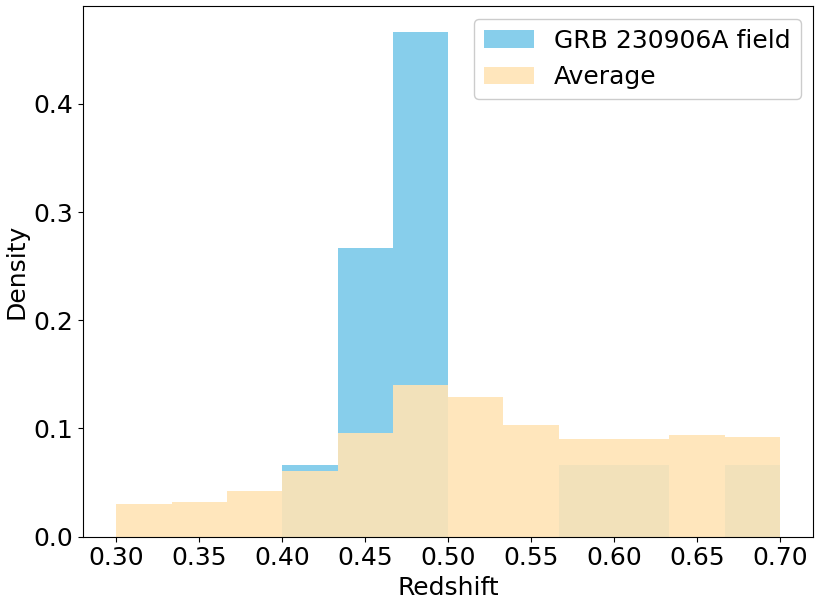}\\
\vspace{0.5cm}
\includegraphics[scale=0.40]{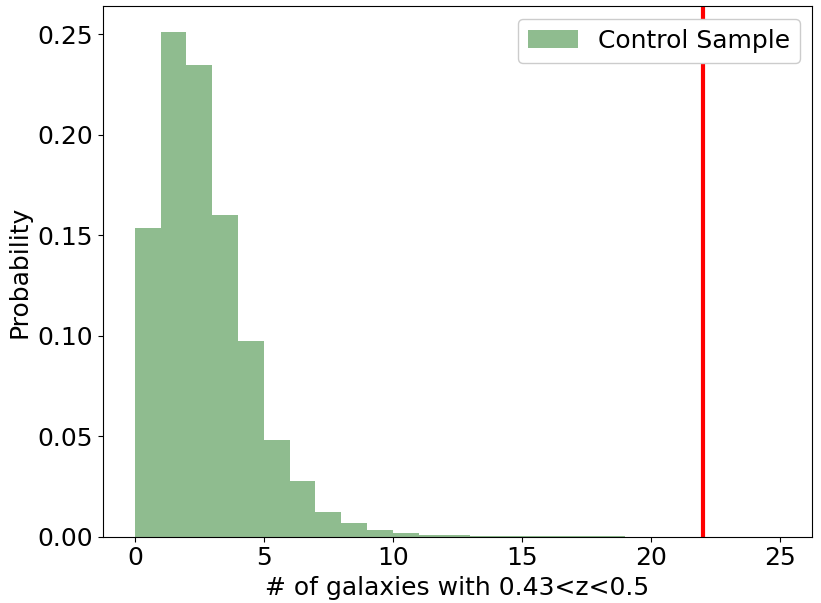}
 \caption{Top: Distribution of the DESI Legacy Survey DR10 photmetric redshift for the field galaxies (within 1 arcmin from the GRB position). 
 The figure is limited to the 0.3 $\leq$ z $\leq$ 0.7 range. Bottom: Probability distribution for the number of galaxies between photo-z 0.43 and 0.5. The red line indicate the galaxies in this range of redshift around the location of GRB 230906A
}
\label{fig:hist}
\end{figure}

Although the spatial coincidence between G$^*$ and GRB 230906A provides tantalizing evidence for a physical association, the probability of a chance coincidence, $P_{\mathrm{cc}} \approx 4\%$, is not negligible. Therefore, we examined the surrounding large-scale environment for other potential host candidates. 
We computed $P_{\mathrm{cc}}$ for all field galaxies  and list only the top ranked in Table~\ref{tab:gal}.  
The derived probabilities are in all cases greater than 0.1, and therefore less favorable than the association with G$^*$.
The next most probable galaxy is marked as G1 with $P_{\mathrm{cc}} \approx$14\%. Its projected angular offset is 
21.7 arcsec, corresponding to a large physical projected offset of $\approx$130 kpc.

\begin{figure*}
\centering
\includegraphics[scale=0.27]{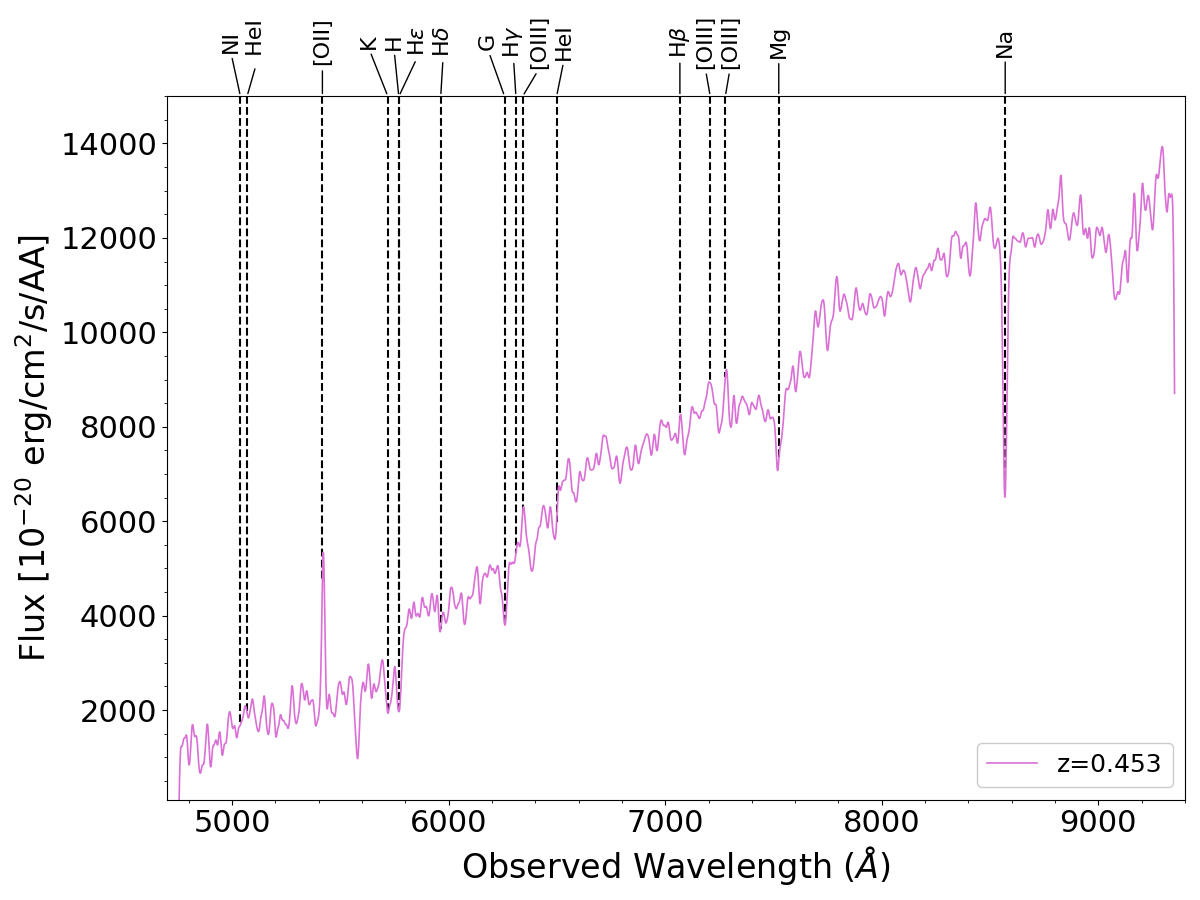}
\includegraphics[scale=0.27]{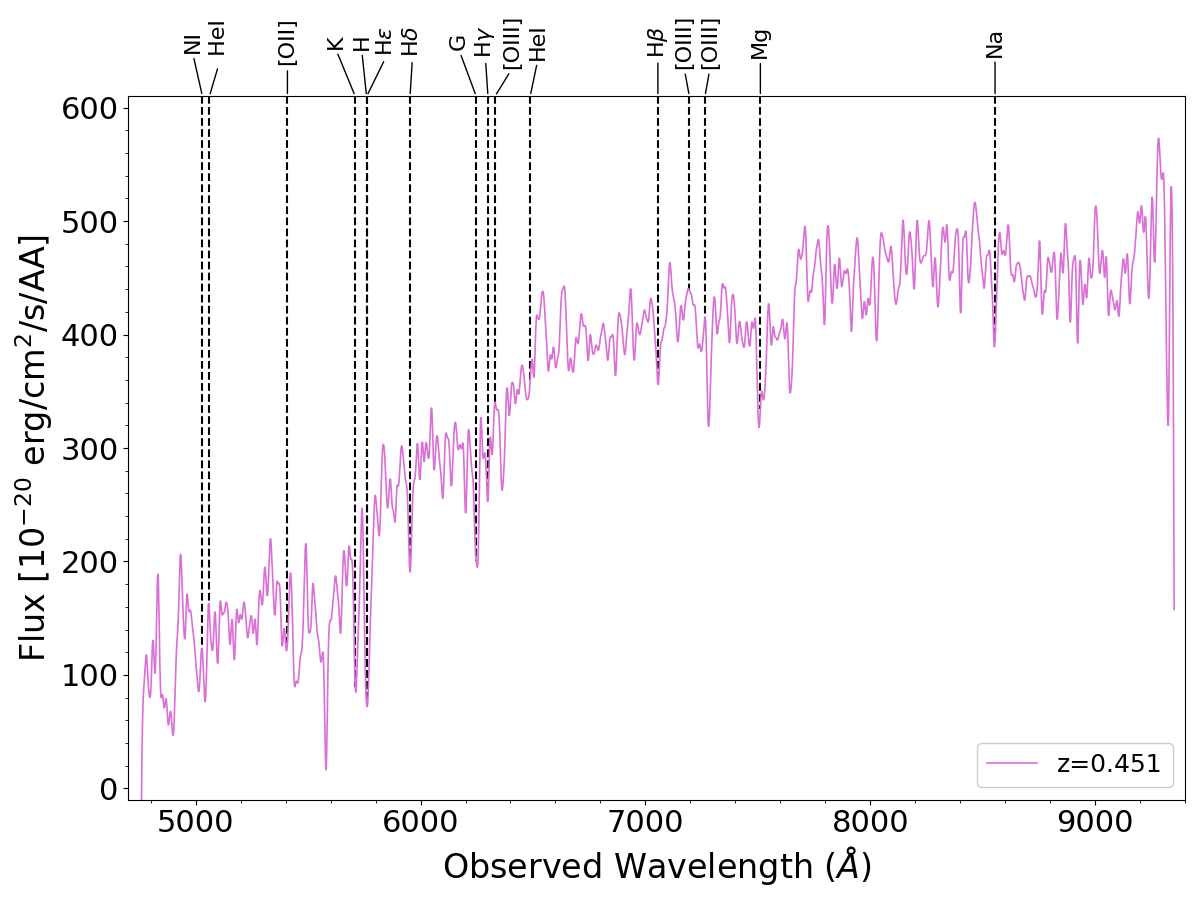}
\includegraphics[scale=0.27]{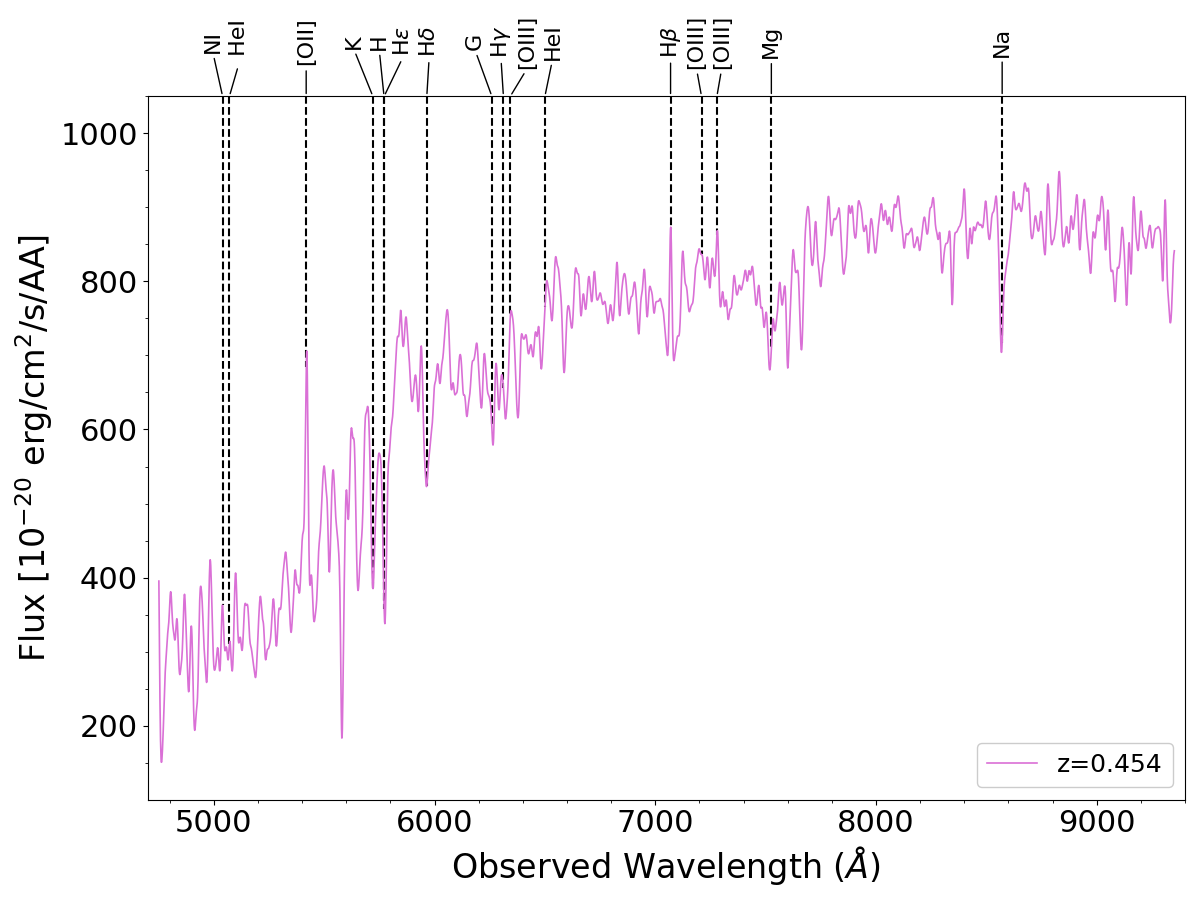}
\includegraphics[scale=0.27]{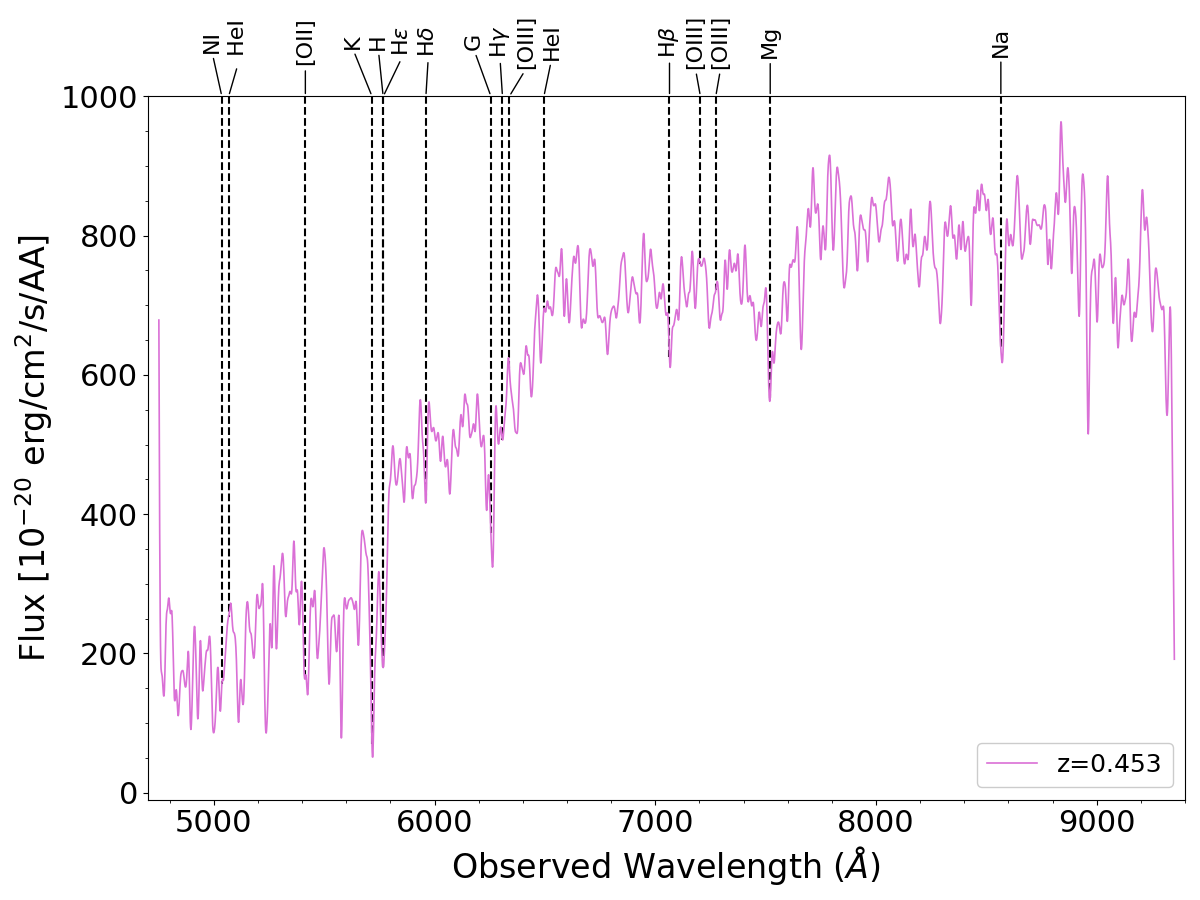}
 \caption{VLT MUSE spectra of 4 galaxies in the field of GRB 230906A. The top-left spectrum is the one extracted for the big galaxy G1 (see Table~\ref{tab:gal}). The spectra has been smoothed with a Gaussian kernel using a FWHM of 10 pixels. 
}
\label{fig:spec}
\end{figure*}

\subsubsection{Association with a galaxy group at $z$=0.453}

Our analysis of the GRB field revealed that several galaxies with similar colors lie close to the GRB position. 
To investigate the possible presence of an overdensity of galaxies at a similar redshift, we studied their distribution of photometric redshifts (photo-$z$) retrieved from DESI Legacy Survey catalog \citep[DR10\footnote{https://www.legacysurvey.org/dr10/};][]{Dey2019}.  Figure~\ref{fig:hist} (top panel) shows the photo-$z$ distribution of all galaxies within a $1\,$arcmin radius of the \textit{Chandra} position, compared with the distribution obtained from a control sample of random sky positions. The control distribution was derived from a set of $\sim$12 million galaxies drawn from the Legacy Survey \citep{2021MNRAS.501.3309Z}. 
From Figure~\ref{fig:hist} we identify a possible excess of galaxies with photo-$z$ between 0.43 and 0.50:  
22 galaxies fall in this redshift bin, 
whereas on average we would expect less than 5 (bottom panel). 

Our MUSE observation covers 1/4th of the search region and includes seven of these galaxies with 
0.43$\lesssim$ photo-z $\lesssim$0.5. Spectroscopic observations confirm that six lie at $\approx$0.45 (see Figure~\ref{fig:spec}) and reveal that the bright central galaxy G1  with $z=$0.453 (rather than the photometric value of 0.58) also belongs to this group of galaxies. 
From the spectroscopic redshift distribution of the candidate group members, we derive a velocity dispersion $\sigma_v$$\approx$400 km s$^{-1}$ and, under the assumption of virial equilibrium, a total dynamical mass of 3$\times$10$^{13}$ $M_{\odot}$ within 400 kpc from the central galaxy G1.
The measured dispersion and the presence of a dominant central galaxy G1 suggest that the system is bound, although the small number of confirmed members and the limited spatial coverage do not allow us to robustly assess its dynamical state.
The derived mass places it in the range of galaxy groups rather than rich clusters such as the one associated with GRB 050509B \citep{Gehrels05,Pedersen05}. It is thus not surprising that our  \textit{Chandra} and \textit{XMM-Newton} imaging did not reveal any diffuse X-ray emission.

The identification of a galaxy group within a GRB field is not a common occurrence. Beside the case of GRB 050509B, only a few more short GRBs were spatially coincident with a cluster or group environment \citep{Prochaska2006,Nugent2020}, although in none of these cases a physical association could be robustly established. Assuming that all possible associations discussed in the literature are a chance alignment, we derive an upper limit $P_{cc}\approx$ 3\% on the probability of a random coincidence between a short GRB and a galaxy group/cluster. 
This is comparable to the probability of random alignment with $G^*$. 

A similar value of $P_{cc}$ is obtained by considering the sky density of galaxy groups. 
The Galaxy and Mass Assembly (GAMA) galaxy group catalogue \citep[G3Cv1;][]{Robotham2011} lists 6,391 groups in the G12 region, which spans 48~deg$^2$. This corresponds to a surface density of roughly 130 galaxy groups per square degree. From this, we infer that the probability of a random GRB sightline intersecting the central core ($\lesssim$1 arcmin) of a galaxy group is $\approx$4\%, consistent with our previous considerations. 

Based on positional arguments, the association between GRB 230906A and the galaxy group at $z$=0.453 is plausible. 
If this is the case, the faint galaxy G$^*$ could be an unrelated background galaxy
and its spatial coincidence with the GRB would be merely a chance alignment. 
In this scenario, GRB 230906A occurred in the intragroup medium, having been ejected from its parent galaxy due to an intrinsic kick velocity and /or 
the effects of tidal interactions. 
However, both of these possibilities appear unlikely.  
The physical offset between GRB 230906A and the nearest member of the group is $\sim 44$\,kpc, which would require extreme conditions in terms of the underlying binary NS  formation kick and its GW time-delay to merger and is inconsistent with findings in other sGRBs, which suggest that extended stellar distributions rather than kicks are the typical origin of large observed offsets \citep{PB2021}.
The alternative possibility that the BNS formed or was perturbed directly via the galaxy merger is also disfavored. This is because the cross section for such an encounter is $\lesssim {\rm AU}^2$, and even then, fine-tuned conditions are required to sent the binary into an orbit with a $\sim 0.5-1$\,Gyr merger time \citep{BP2024}. 
Moreover, our study of the GRB prompt and afterglow phases shows that the X-ray to gamma-ray flux ratio is rather typical for a short GRB and does not point to a low-density environment as expected for an explosion in the intergalactic medium \citep{Mulchaey00}.  For example, GRB 211211A and GRB 230307A were localized far away from their galaxies
and displayed a much lower flux ratio log($F_{X,11}$/$f_{\gamma,15-150}$) $\approx-8$ \citep{Troja22,Yang24}, indicative of a low-density circumburst medium \citep{oconnor2020,oconnor2022}. 

For all these reasons, we disfavor the hypothesis of a chance alignment and consider the alternative possibility that G$^*$  is indeed a group member at $z\!\approx\!0.45$.
If G$^*$ is associated with the galaxy group, then the coincidence probability of G$^*$ being located at the position of the GRB drops significantly, indicating that the GRB occurred within G$^*$.

The measured flux would imply a mass of $\approx6\times 10^7 M_{\odot}$ consistent with those of tidal dwarf galaxies \citep{Hunsberger1996}.
At this redshift, the properties of G$^*$ (its faintness and relatively blue color, $r$-$H\approx$0.3 mag) would place it well outside of the typical short GRB host population (Figure~\ref{fig:galaxies}). 
Since most of the comparison hosts are field galaxies, they may not provide reliable guidance in dense environments, where interactions among galaxies influence their morphology and stellar population \citep{Toomre72,Moore96,Poggianti25}. 

\begin{figure}
\centering
\includegraphics[scale=0.31]{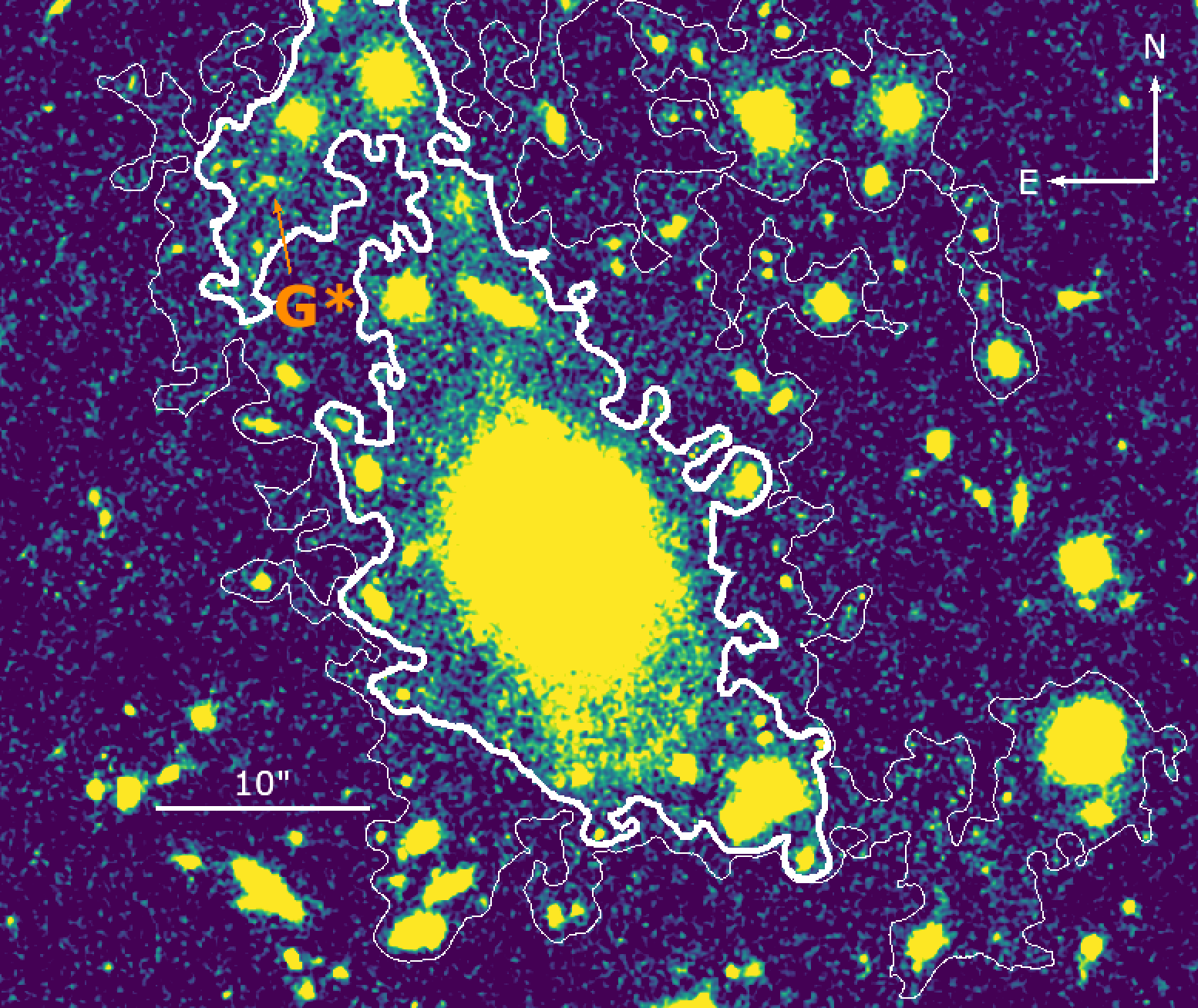}
 \caption{Contours derived using the HST $F160W$ image and considering 1 DN (white thin line) and 10 DNs (white thick line) above the background. 
}
\label{fig:tail}
\end{figure}

Evidence for recent galaxy mergers was sought via inspection of tidal features, obtained by tracing \textsc{DS9} \citep{DS9} isophotal contours on the smoothed
\textit{HST} image. Following \citet{Mullan2011} and \citet{Knierman2003}, 
we defined an ``in-tail" region a contiguous regions with 1 data number (DN) count above the background, corresponding to a brightness of $\approx$27.5 mag arcsec$^{-2}$.  In Figure~\ref{fig:tail} we show that the location of GRB 230906A and G$^{*}$ falls within the ``in-tail'' region, which
extends for about 180 kpc north-east and south-west of G1, and appears slightly curved at its tips. Other knots of emission are visible close to G$^{*}$, their brightness and morphology is reminiscent of the tidal dwarfs presented in \citet{Hunsberger1996}. 

The morphology of the tail suggests an encounter with a spiral galaxy, and its long extent points to an old interaction age. 
Adopting a typical disk circular speed of 250\,km\,s$^{-1}$ as an estimate of the debris streaming velocity, the tail’s length implies a formation timescale of $\tau \approx L/v\approx$700 Myr. This timescale is typical of merger delay times \citep{BP2019}. This suggests a natural origin for the binary massive star progenitor of GRB 230906A, due to enhanced star formation induced by the galaxy merger. This, after $\sim 0.5$-$1$\,Gyr, resulted in a NS merger and a short GRB. 

A binary NS merger within the tidal tail has interesting implications for $r$-process enrichment. The narrow width, low gas density and velocity dispersion of such tails mean that the ejecta can expand outwards up to a large distance 
$r_{\rm fade}\approx 460 (\sigma_v/20\mbox{km s}^{-1})^{-0.4}(E/10^{51}\mbox{erg})^{0.28}\allowbreak
(n/10^{-3}\mbox{ cm}^{-3})^{-0.365}$ 
before it can start mixing efficiently with the interstellar gas \citep{Beniamini2018}.

For typical parameters of tidal tails \citep{DucRenaud2013}, this can be a significant fraction of their width, and lead to a large fraction of the $r$-process synthesized mass being lost to the circumgalactic medium (CGM). 
Therefore, events such as these, despite their rarity, offer a natural pathway to elevated halo-phase r-process abundances in interacting systems and groups, consistent with the broader picture in which the CGM acts as a major reservoir of metals \citep{Tumlinson2017}.

\begin{figure}
\centering
\includegraphics[scale=0.37]{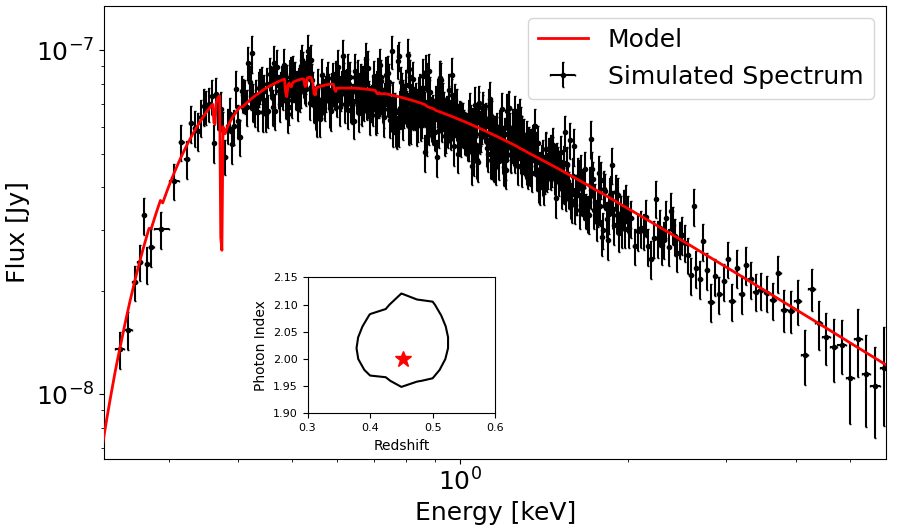}
 \caption{Simulated \textit{NewAthena} X-IFU afterglow spectrum for an integrated flux of $5 \times 10^{-13}$ erg cm$^{-2}$ s$^{-1}$ (0.3–10 keV). The absorbed power-law model is overplotted in red. The inset shows the contour plot for redshift and photon index. The red star indicates the values used for the simulated spectrum and solid lines represent the 90\% confidence level.
}
\label{fig:athena}
\end{figure}

\subsubsection{Breaking the degeneracy with X-ray spectroscopy}

The case of GRB 230906A highlights the challenges in determining the true GRB distance scale when relying solely on positional arguments and a posteriori associations with nearby galaxies.

In this instance, a sub-arcsecond localization was crucial to establish its likely connection with the faint galaxy G$^*$, an association that would have been dismissed otherwise: an XRT localization with typical error radius of $\sim$2 arcsec would yield $P_{cc}>50\%$ and favor the brighter galaxy G1 as putative host. Yet, the complex environment of GRB 230906A and the extreme faintness of G$^*$ prevented us from obtaining the GRB redshift. At present, such a measurement would be indirectly attainable only with deep \textit{JWST} observations of its putative host galaxy. 

The most robust method to establish a GRB distance remains rapid afterglow spectroscopy, typically conducted with ground-based optical or near-infrared facilities. However, the intrinsically faint and rapidly fading optical afterglows of short GRBs make this approach rarely successful: fewer than 10\% of short GRB redshifts are obtained from afterglow spectra \citep{DeUgartePostigo13, oconnor2022, AguiFernandez2023}. As discussed in Sect.~\ref{sec:discussion}, the optical afterglow of GRB 230906A could have been detected with deeper, earlier observations, but would still have been too faint for meaningful spectroscopy.

High-resolution X-ray spectroscopy, pioneered by the gratings on \textit{XMM-Newton} and \textit{Chandra} \citep{denHerder2001,Canizares2005}, 
has so far been restricted to the brightest sources. A new era
has recently begun with the X-Ray Imaging and Spectroscopy Mission 
\citep[XRISM;][]{XRISM}, whose sensitivity and limited 
ToO capabilities are not optimal for GRB afterglows. In the next decade, the \textit{NewAthena} mission will deliver order-of-magnitude gains in sensitivity, angular resolution, and ToO response time \citep{Cruise2024}, opening the door to systematic high-resolution X-ray spectroscopy of GRB afterglows, including those from short bursts.

At a redshift $z\sim$0.453, prominent absorption edges, such as the neutral oxygen K edge and the iron L complex, and multiple resonance absorption lines fall within the \textit{NewAthena}/X-IFU bandpass and can be discriminated from Galactic structures thanks to X-IFU's resolution \citep{Barret2023}. To illustrate this, in Figure~\ref{fig:athena} we present a simulated GRB afterglow spectrum derived with the Simulation of X-ray Telescopes software v.3.2.0 (\textsc{SIXTE}; \citealt{Dauser19}). 
The simulation was ran assuming an X-ray flux of $5\times$10$^{-13}$ erg\,cm$^{-2}$\,s$^{-1}$, 
an intrinsic absorption of $N_{H,z} \approx$10$^{21}$ cm$^{-2}$ and a total exposure of 50 ks.
The photon index was fixed to $\Gamma$=2.0 and the Galactic neutral hydrogen column was set to $N_H \approx$\,$3\times10^{20}$ cm$^{-2}$.
For the instrumental parameters, we adopted a mirror of 13 rows, a spectral resolution of 4 eV, and set the non-X-ray background (NXB) to $5\times10^{-3}$ counts cm$^{-2}$ s$^{-1}$ keV$^{-1}$ in the 2-10 keV range.
The simulated spectrum was then fit with an absorbed power-law model
(\texttt{tbabs*ztbabs*pegpwrlw}) in \textsc{XSPEC} \citep{Arnaud96}, with the redshift treated as a free parameter. The resulting uncertainties, shown in the inset of Figure~\ref{fig:athena}, demonstrate that the true redshift can be constrained even for relatively faint fluxes.

\subsection{Burst energetics and SN limits}

Although the distance scale of GRB 230906A is not unambiguously determined, the study of its environment allows us to set a robust lower bound of $z\gtrsim$0.45. At such a distance, any kilonova emission would be too faint to detect with ground-based observatories. Consequently, the properties of the GRB progenitor and central engine must be inferred through indirect diagnostics, such as the burst’s energy output \citep{Burrows05,Cenko09,Troja2016} and the lack of a bright SN \citep{Bloom06,Berger13}. 

In the following, we consider three possible distance scales for GRB 230906A: $z=0.45$ set by its possible association with the group of galaxies,  $z=1.2$ derived by assuming that its putative host galaxy G$^*$ fits within the typical luminosity of short GRB galaxies, and $z=3$ derived by assuming that the Balmer break of G$^*$ lies beyond the $F160W$ band.  
Using the best fit spectral model described in Sect.~\ref{sec:GBM}, we derive an isotropic equivalent gamma-ray energy 
$E_{\gamma,{\rm iso}}$\,$\approx$\,2$\times$10$^{51}$ erg, 1.4$\times$10$^{52}$ erg, 
and 6.8$\times$10$^{52}$ erg for $z$=0.45, $z$=1.2 and $z$=3, respectively (1-10000 keV, rest frame). 
Assuming a typical gamma-ray efficiency $\eta_{\gamma}$ $\sim$ 15\% \citep{Beniamini16},  we infer the isotropic-equivalent kinetic energy of the blast-wave as $E_{K,{\rm iso}}$\,=\,$E_{\gamma,{\rm iso}}$($\eta_{\gamma}^{-1}$ - 1)$\,\approx$\,6\,$E_{\gamma,{\rm iso}}$,   and the total energy of the explosion as $E_{{\rm iso}}$\,=\,$E_{\gamma,{\rm iso}} + E_{K,{\rm iso}}$. 

Using the limit on the jet-break time, $t_j \gtrsim 6$~d (90\% confidence level), derived from the \textit{XMM-Newton} observation (Sect.~\ref{sec:XRAY}), we set a lower limit on the jet opening angle \citep{sari1999,vaneerten12}:

\begin{equation}
\begin{split}
\theta_{j} &= 3.6^{\circ} \left(\frac{t_{j}}{6\,{\rm d}}\right)^{3/8}
\left(\frac{1+z}{2}\right)^{-3/8} \\
&\quad \times \left(\frac{E_{K,{\rm iso}}}{10^{53}\,{\rm erg}}\right)^{-1/8}
\left(\frac{n}{10^{-3}\,{\rm cm^{-3}}}\right)^{1/8}
\end{split}
\end{equation}

where we assumed a top-hat jet profile and an on-axis orientation. 
For a density $n\approx10^{-3}\,{\rm cm^{-3}}$, we find $\theta_{j}\gtrsim$6.7$^{\circ}$, $\theta_{j}\gtrsim$4.5$^{\circ}$ and $\theta_{j}\gtrsim$2.9$^{\circ}$ for $z$=0.45, $z$=1.2, and $z$=3, respectively. 
From the above constraints, we derive a collimation-corrected total energy $E \gtrsim$ $\left(0.8-5\right) \times$10$^{50}$ erg, which is well within the energy budget of NS mergers.

If GRB 230906A was powered by a standard accreting black hole (BH), the released energy relates to the accreted torus mass as $E \approx \eta M_{\rm acc} c^2$, where $\eta$ is the efficiency in converting the accreted torus mass into jet energy and is sensitively dependent on the energy extraction mechanism.  A neutrino cooled disk with $\eta \approx$10$^{-4}$-10$^{-2}$ \citep{Popham1999,DiMatteo2002,ChenBeloborodov2006,ZalameaBeloborodov2011,LiuGuZhang2017}
would thus be rather massive, $M_{\rm acc} > 5 \times 10^{-3}$ $M_{\odot}$,
and the short GRB duration ($T_{90}\approx0.9$\,s) would imply correspondingly high accretion rates. 
Moreover, only a fraction of the initial torus mass actually accretes onto the BH as the rest is expelled in powerful winds. Recent observations of kilonovae often show an early blue component likely powered by this wind ejecta
\citep{Troja23, Rastinejad2023, kn070809, Troja18}, suggesting that disc outflows are frequent and efficiently remove mass from the torus \citep{Janiuk19}. The post-merger torus mass is thus  $M_{\rm torus}\approx M_{wind} + M_{\rm acc}$, 
where, for reference, the wind ejecta mass inferred from the kilonova AT2017gfo was $M_{wind}\approx$0.01 $M_{\odot}$ \citep{Kasen17,Kedia23,Ristic25}. 
This further tightens the requirements on the NS merger model, showing that neutrino-antineutrino ($\nu\bar\nu$) annihilation would be viable only if $\eta$ is near its maximum value. 
Instead, magnetically driven mechanisms \citep[e.g.][]{BZ77,Narayan2003,Rezzolla2011,Paschalidis2015,Ruiz2016} allow for substantially higher efficiencies and can naturally reproduce the GRB energy output even for low-mass tori, $M_{\rm torus}\approx 10^{-4}$-$10^{-2} M_{\odot}$.  
In agreement with our conclusions, \citet{Wu2025} recently suggested that a BZ jet better explains the data of short GRBs and that it implies a jet formation efficiency of $\sim 10^{-3}$-$10^{-2}$.

\begin{figure}
\centering
\includegraphics[width=0.45\textwidth]{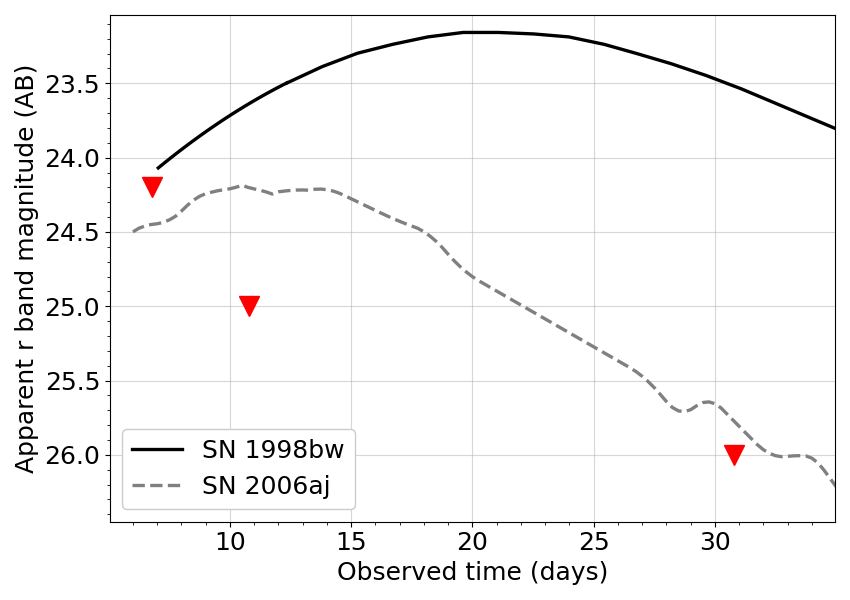}
 \caption{Optical limits for GRB 230906A compared to the lightcurves of two proto-typical GRB-SNe, SN1998bw (solid line; \citealt{Galama98,Clocchiatti2011}) and SN2006aj (dashed line; \citealt{Pian06,Brown2014}), redshifted at $z=$0.453.}
 \label{fig:sn}
\end{figure}

We conclude showing that if GRB 230906A is associated with the group of galaxies at $z$\,$\sim$\,0.453, then the available optical limits can be used to constrain the presence of an underlying SN \citep{Galama98,Bloom99}. 
This comparison, which strengthens the identification of GRB 230906A as a merger-driven GRB, is presented in Figure~\ref{fig:sn}. 
For the higher redshift solution, the presence of a SN cannot be probed by our observations. Nonetheless, the short $T_{90}$ implies a high probability of a non-collapsar burst, $f_{NC}$\,$\approx$\,0.7 \citep{Bromberg13}. 

\section{Summary and Conclusions}
\label{sec:summary}

We presented a detailed analysis of the short duration ($T_{90}$\,$\sim$\,0.9~s) GRB 230906A and its environment. 
The burst was discovered by the \textit{Fermi} GBM and triangulated by the IPN to a narrow error region, later refined by \textit{Swift} thanks to the identification of the burst X-ray counterpart. 
Thus, this is one of the few well-localized short GRBs not initially discovered by the \textit{Swift} Burst Alert Telescope \citep{BAT2005}. 

We obtained a sub-arcsecond \textit{Chandra} localization and deep \textit{HST} imaging uncovered a very faint source ($G^{*}$; $F160W\!\sim\!26$ AB) at the afterglow position.
We consider $G^{*}$ as the putative host galaxy, although its faintness places it at the bottom of the distribution of short GRB hosts. 
We discuss two possible scenarios:

$\bullet$ a high-redshift origin would explain the faintness and compactness of $G^{*}$. To be consistent with the typical brightness of short GRB galaxies, it should lie at $z\gtrsim$1.2. 
An even higher redshift, $z\gtrsim$3, would easily explain the flat SED as the Balmer break lies beyond the $F160W$ band.
If confirmed through e.g. JWST spectroscopy, GRB 230906A would be one of the highest redshift short GRBs known to date. 

$\bullet$ a galaxy group environment where faint satellites and tidally formed dwarfs are common. 
In our images we identify an overdensity of galaxies along the GRB sighline. Integral-field spectroscopy with VLT/MUSE confirms this as a group at $z\approx0.453$, with a velocity dispersion $\sim$\,400\,km\,s$^{-1}$ and a dynamical mass $\sim$\,$3\times10^{13}\,M_\odot$. 
\textit{HST} images reveal an extended tidal debris around the central group galaxy G1, indicative of an old galaxy–galaxy interaction.
Whereas there is a negligible probability associated with a NS merger induced directly by the galaxy merger, an indirect path, whereby the galaxy merger enhances star formation eventually leading to the NS merger, is reasonably plausible.\\

GRB~230906A demonstrates that, in the absence of an optical localization, an accurate X-ray position is crucial to identify the candidate GRB host galaxy that would otherwise be missed or misassigned.
This accuracy is delivered only by \textit{Chandra}. 

Our analysis of the field places a conservative lower bound of $z\gtrsim$ 0.45 on the burst distance scale, 
although to rule out the high redshift ($z\gtrsim$1.2) scenario, additional observations with \textit{JWST} would be required. 

In future events, the lingering ambiguity in redshift assignment may be resolved by next-generation X-ray capabilities. High-resolution X-ray spectroscopy on fast ToOs, such as those enabled by \textit{NewAthena}/X-IFU, can measure absorption edges and resonance lines directly in the X-ray afterglow, unambiguously breaking the redshift degeneracy when optical spectroscopy is impractical.

\section*{ACKNOWLEDGEMENTS}

The authors would like to thank Kim Page and Phil Evans for their help in refining the analysis of the XRT data and for useful discussions. They also thank Caryl Gronwall for the insightful discussion regarding the possible association of this event with a galaxy merger within the group.

The material is based upon work supported by National Aeronautics and Space Administration under award
number 80NSSC22K1516 and by the Smithsonian Astrophysical Observatory under award GO3-24047X.
E.T. and Y.-H.Y. were supported by the European Research Council through the Consolidator grant BHianca (grant agreement ID 101002761).
Part of this work was carried out at the Aspen Center for Physics, which is supported by National Science Foundation grant PHY-2210452. B.O. gratefully acknowledges support from the McWilliams Postdoctoral Fellowship in the McWilliams Center for Cosmology and Astrophysics at Carnegie Mellon University. D.T. and D.N.'s research was supported by UK Science and Technology Facilities Council grant ST/X001067/1 and Royal Society research grant RGS /\ R1 /\ 231499.

The scientific results reported in this article are based on observations made by the Chandra X-ray Observatory (CXO) under Program 24500202 (PI: Dichiara) and XMM-Newton under the Program 08830 (PI:Troja).

This paper employs a list of Chandra datasets, obtained by the Chandra X-ray Observatory, contained in the Chandra Data Collection (CDC) ~\dataset[DOI: 10.25574/cdc.510]{https://doi.org/10.25574/cdc.510}. This research has made use of software provided by the Chandra X-ray Center (CXC) in the application package \texttt{CIAO}.

The HST data presented in this article were obtained from the Mikulski Archive for Space Telescopes (MAST) at the Space Telescope Science Institute. The specific observations analyzed can be accessed via \dataset[DOI: 10.17909/xded-jp95]{https://doi.org/10.17909/xded-jp95}.

This work was also based on observations with the NASA/ESA Hubble Space Telescope obtained from the Mikulski Archive for Space Telescopes (MAST) hosted at the Space Telescope Science Institute, which is operated by the Association of Universities for Research in Astronomy, Incorporated, under NASA contract NAS5-26555.  

This research has made use of data and software provided by the High Energy Astrophysics Science Archive Research Center (HEASARC), which is a service of the Astrophysics Science Division at NASA/GSFC and the High Energy Astrophysics Division of the Smithsonian Astrophysical Observatory. This work made use of data supplied by the UK \textit{Swift} Science Data Centre at the University of Leicester. In addition, this research utilized the XRT Data Analysis Software (XRTDAS), developed under the responsibility of the ASI Science Data Center (ASDC), Italy.

In this work we used data collected at the European Southern Observatory under ESO programs 110.24CF and 114.27LW.

\vspace{5mm}
\facilities{Fermi, Swift, CXO, XMM-Newton, HST, VLT}


\software{Fermitools \citep{2019ascl.soft05011F},
          RMFIT \citep{2014ascl.soft09011G},
          ligo.skymap \citep{2016PhRvD..93b4013S},
          CIAO \citep{Fruscione2006},
          SAS,
          XSPEC \citep{Arnaud96},
          emcee \citep{emcee},
          SCAMP\citep{SCAMP},
          SWarp \citep{Swarp},
          HOTPANTS \citep{Becker2015}, 
          SFFT \citep{SFFT},
          DrizzlePac \citep{Gonzaga12},
          Source Extractor \citep{Bertin96},
          ESO-Reflex \citep{2013A&A...559A..96F},
          IRAF \citep{1986SPIE..627..733T},
          DS9 \citep{DS9},
          SIXTE \citep{Dauser19}
          }









\bibliography{sample631}{}
\bibliographystyle{aasjournal}


\appendix

\section{Spectroscopic redshift of field galaxies}

The complete list of galaxies with spectroscopic redshifts derived from MUSE observations is presented in the following Table~\ref{tab:musespec}. A redshift was successfully measured for 18 galaxies in the field of GRB 230906A, six of which share a common redshift of z$\sim$0.45.

\begin{table}[b!]
 	\centering
 	\caption{Spectroscopic redshift value derived from MUSE observations of 230906A field galaxies
 	}
 	\label{tab:musespec}
 	\begin{tabular}{llcc}
    \hline
   RA  & Dec. & spec-z & photo-z$^{a}$\\
 \hline
79.7507 & -47.8970$^{G1}$  & 0.453 &  $0.58 \pm 0.01$\\
79.7516 & -47.8963$^{G6}$ &  0.451 & $0.4 \pm 0.1$\\
79.7519 & -47.8938$^{G5}$ &  0.438 & $0.48 \pm 0.09$\\
79.7538 & -47.8937$^{G3}$ &  0.737 & $0.6 \pm 0.3$\\
79.7540 & -47.8909$^{G2}$ &  0.453 & $0.49 \pm 0.05$\\    
79.7543 & -47.8892$^{G7}$ &  0.852 & $0.74 \pm 0.09$\\
79.7467 & -47.8915  &     0.302 & $0.48 \pm 0.09$\\
79.7455 & -47.8938  &     0.465 & $0.49 \pm 0.08$\\
79.7406 & -47.8972  &     0.456 & $0.44 \pm 0.04$\\
79.7406 & -47.8995  &     0.302 & $0.35 \pm 0.07$  \\
79.7467 & -47.9001  &     0.453 & $0.50 \pm 0.04$\\
79.7481 & -47.9032  &     0.454 & $0.40 \pm 0.05$\\
79.7436 & -47.9021  &     0.561 & $0.49 \pm 0.07$\\
79.7397 & -47.8882  &     0.687 & $0.7 \pm 0.1$\\
79.7525 & -47.8879  &     0.682 & $0.6 \pm 0.3$\\
79.7535 & -47.9020  &     0.416 & $0.5 \pm 0.1$ \\
79.7473 & -47.9006  &     0.919 & $1.1 \pm 0.1$\\
79.7565 & -47.9015  &     0.892 & $1.3 \pm 0.1$\\
 \hline
 \hline
    \end{tabular}
\begin{flushleft}
    \quad \footnotesize{$^{a}$ Z\_PHOT\_MEAN and Z\_PHOT\_STD retrieved from DESI Legacy Survey DR10. }\\
\end{flushleft}
\end{table}

\label{lastpage}

\end{document}